# Theory of Pixel Lensing

**Andrew Gould**[1]

Dept of Astronomy, Ohio State University, Columbus, OH 43210

e-mail gould@payne.mps.ohio-state.edu

**Abstract**

Pixel lensing, gravitational microlensing of unresolved stars, is potentially much more sensitive and much more widely applicable than is generally recognized. I give explicit expressions for the pixel noise induced by a time-variable PSF, by photometric and geometric misalignment, and by discrete pixelization, and I show that these can all be reduced below the photon noise. Pixel lensing can be divided into two regimes. In the "semi-classical" regime, it is similar to classical lensing in that it measures the time scale of individual events. In the "spike" regime, it measures the total optical depth but not individual time scales. I present simple expressions for the boundary between the two regimes and for the event rate in the latter one. These expressions can be used to quickly classify all potential pixel lensing experiments. Pixel lensing can measure the luminosity function as well as the mass function of stars in target galaxies to a distance of a few Mpc. Future space-based pixel lensing could be $\sim 5$ times more effective in the infrared than in the optical, depending on developments in detector technology. Pixel techniques can also be applied to non-pixel-lensing problems including the search for unresolved variable stars and follow up observations of lensing events found in classical lensing searches. To benefit fully from pixel-lensing techniques, follow-up observations should have resolutions of at least 5 pixels per FWHM.

Subject Headings: gravitational lensing – stars: masses – stars: luminosities – techniques: photometric

# 1. Introduction

Pixel lensing is gravitational microlensing of unresolved stars. In classical lensing surveys of the type proposed by Paczyński (1986), one monitors large numbers of resolved stars in hopes of finding a few which are microlensed by foreground objects. The magnification $A(x)$ depends only on the ratio $x = \theta/\theta_e$ of the angular separation of the source and the lens to the Einstein radius,

$$A(x) = \frac{x^2 + 2}{x(x^2 + 4)^{1/2}}, \qquad x \equiv \frac{\theta}{\theta_e} \tag{1.1}$$

where

$$\theta_e^2 \equiv \frac{4GM d_{\rm ol} d_{\rm ls}}{c^2 d_{\rm os}}, \tag{1.2}$$

$M$ is the mass of the lens, and $d_{\rm ol}$, $d_{\rm ls}$, and $d_{\rm os}$ are the distances between the observer, source, and lens. At least four such surveys are currently being carried out toward the Large and Small Magellanic Clouds (LMC and SMC) and the Galactic bulge and at least 100 candidate microlensing events have been reported to date (Alcock et al. 1993, 1995a,b; Aubourg et al. 1993; Udalski et al. 1994a; Allard 1995). The motion of the lens relative to the observer-source line of sight is usually well-approximated as a uniform velocity, $\mathbf{v}$, in which case the separation is a function of only three parameters,

$$x(t; t_0, \beta, \omega) = [\omega^2(t - t_0)^2 + \beta^2]^{1/2}, \qquad \omega \equiv \frac{v}{d_{\rm ol}\theta_e}, \tag{1.3}$$

where $t_0$ is the time of maximum magnification, $\beta$ is the impact parameter in units of $\theta_e$, and $\omega^{-1}$ is the Einstein crossing time. The optical depth $\tau$ can then be found directly from the observed time scales $\omega_i^{-1}$,

$$\tau = \frac{\pi}{2NT} \sum_i \frac{1}{\omega_i \epsilon_i}, \tag{1.4}$$

where $N$ is the total number of stars observed, $T$ is the duration of the observations, and $\epsilon_i$ are the efficiencies of detection.

Note that the event rate $\Gamma_0$ per observed star is a function not only of $\tau$ but also of the typical time scale,

$$\Gamma_0 = \frac{2}{\pi}\omega\tau. \tag{1.5}$$

While classical lensing searches are a fantastic success, they are fundamentally limited to galaxies with substantial numbers of resolved stars, namely the three galaxies currently being searched. There is a great deal that one could hope to learn from lensing surveys of other galaxies. For example, the halo of M31 could be probed for lenses by monitoring stars in the disk of M31 (Crotts 1992; Baillon et al. 1993). Because of the M31's high inclination, the lensing rate should be much higher toward the far side of the disk compared to the near side (Crotts 1992), which would be an unambiguous signature of lensing. One could measure the mass function of the bulge of M31 by monitoring the stars in the M31 bulge. One could look for intra-cluster lenses in the Virgo cluster by monitoring M87 (Gould 1995b), or even study the star-formation history of the universe by searching for lenses at cosmological distances (Gould 1995a). All of these projects require pixel lensing, the detection of lensing of unresolved stars.

In this paper, I present a general theory of pixel lensing. I show that there are two distinct regimes of pixel lensing which I label "semi-classical" and "spike". The semi-classical regime is similar to classical lensing. Indeed for nearby galaxies like the LMC, semi-classical pixel lensing may be regarded as a simple extension of classical lensing beyond the so-called "crowding limit". Moderately distant galaxies like M31 are generally still in the semi-classical regime. Classical lensing in these galaxies is virtually impossible, but pixel lensing yields very much the same type of information as classical lensing does in galaxies that are nearer. In particular it is possible to measure the individual time scales of the events and so use equation (1.4) to determine the optical depth. It is also possible to measure the unlensed fluxes of the source stars and thus the luminosity function (LF). In more distant galaxies, pixel lensing enters the spike regime. It is no longer possible to measure the time scales of individual events nor the unlensed fluxes of the sources. Remarkably, however, it is still possible to measure the optical depth.

I present simple formulae for the boundary between the two regimes and for the number of events which can be detected in the spike regime with a given signal to noise, assuming that the noise is dominated by photon statistics. There are several sources of systematic error which could in principle take precedence over the photon noise. These include noise introduced by variation in the seeing, by problems in aligning the images geometrically and photometrically, and noise from discrete pixelization of the point spread function (PSF). I derive explicit formulae which allow one to determine whether these systematic sources of noise are significant relative to photon noise.

The formalism presented here should prove useful in optimizing the design of pixel lensing experiments, in comparing different possible experiments, and in developing new pixel lensing ideas. I give a few illustrative examples of these applications.

Two groups are currently carrying out pixel lensing searches toward M31,

Crotts and Tomaney (Crotts 1992) and AGAPE (Baillon et al. 1993, Melchior 1995). Both groups have for their initial search adopted a "threshold" approach to identifying events. In this approach, one demands that the event rises above the background by a minimum number of $\sigma$ for several consecutive measurements. Colley (1995) has analyzed the detectability of events toward M31 by simulating this threshold approach.

The framework adopted here is quite different. I analyze detectability from the standpoint of the total signal to noise of the event assuming that photon statistics are the only source of noise and that the entire sequence of images can be searched with a set of optimal event filters. Many will undoubtedly regard these assumptions as symptoms of the wild optimism and hopeless naiveté of a theorist. I was led to this approach after visiting the AGAPE group and being shown the difference between two deep images of M31 taken in similar seeing. The individual images were highly mottled with surface brightness fluctuations, but the difference contained only a single "star" plus photon noise. Subsequently, Tomaney & Crotts (1996) have used convolution to produce differences of images taken in substantially different seeing. These difference images are also near the photon-noise limit. In one, for example, the empirically calibrated flux error is only $R = 24.6$ in a region of M31 with surface brightness $R = 18.2$ mag arcsec$^2$. The practical challenges to pixel lensing are being overcome by the observers. Theory is not wildly out in front of these advances. On the contrary, it is struggling to catch up with them.

## 2. Pixel Lensing Light Curves

In classical lensing, one does not actually measure $A$. Rather one measures the total flux $F$ from the lensed "star",

$$F(t; t_0, \beta, \omega, F_0, B) = F_0 A(t; t_0, \beta, \omega) + B', \qquad (2.1)$$

where $A$ is given by equations (1.1) and (1.3), $F_0$ is the flux from the star in the absence of lensing, and $B'$ is the flux from any unlensed blended source within the star's PSF. If $B'$ is assumed to be zero (as it often is), then $F_0$ can be very well determined from the unlensed portion of the light curve. In this case the measurement of $F$ does directly yield $A$. However, since lensing searches are always carried out in crowded fields (and since in any event the lens itself can contribute unlensed light) the blend term cannot be assumed to be 0. The only quantity that is really known from the unlensed light curve is $B = B' + F_0$. Rewriting equation

(2.1) in terms of $B$ gives,

$$F(t;t_0,\beta,\omega,F_0,B) = F_0[A(t;t_0,\beta,\omega) - 1] + B. \tag{2.2}$$

Since $B$ is very well known, the observed light curve when written in this form must be fit to four unknown parameters, $F_0, t_0, \beta$, and $\omega$.

The basic idea of pixel lensing is to subtract from the current image $I(t)$ a reference image $R$ derived from images taken before or after the event occurs. The actual construction of a reference image is a rather subtle question which will be addressed in § 9.6. For the present, I will assume that $R$ is identical to $I(t)$ except that 1) the star is not lensed and 2) $R$ has much less noise than $I(t)$. The difference image $D(t)$,

$$D(t) = I(t) - R, \tag{2.3}$$

is then a perfectly blank field except for 1) a PSF with total flux $F = F_0(A - 1)$ and 2) photon noise. That is, the flux is given by equation (2.2) with $B = 0$. Since $B$ is a known constant for both pixel lensing and classical lensing, the structures of the light curves are essentially the same for the two cases.

Pixel lensing differs from classical lensing in only one fundamental way. In pixel lensing there are many stars per resolution element, so the photon noise is dominated by light from stars not being lensed. This means that the noise is virtually independent of the magnification. In classical lensing by contrast the photon noise is generally dominated by light from the lensed star. Thus, while the form of the lensing signal given by equation (2.2) is the same, the form of the noise is different.

For spike pixel lensing there is a second difference. In this regime there are so many unresolved stars per resolution element, one is generally able to observe events only when $\beta \ll 1$. The excess flux then takes the limiting form,

$$F_0[A(t;t_0,\beta,\omega) - 1] \to \frac{F_0}{\beta} G(t;t_0,\omega_{\text{eff}}), \tag{2.4}$$

where

$$G(t;t_0,\omega_{\text{eff}}) \equiv \left[\omega_{\text{eff}}^2(t-t_0)^2 + 1\right]^{-1/2}, \qquad \omega_{\text{eff}} \equiv \frac{\omega}{\beta}, \tag{2.5}$$

and $\omega_{\text{eff}}$ is the effective time scale. For reasons which will become clear below, I refer to $G$ as a "filter function". Notice that whereas four parameters are required to describe an ordinary lensing event, only three ($F_0/\beta$, $t_0$, and $\omega_{\text{eff}}$) are required to describe events in the spike regime. That is, to the extent that the light curve is adequately described by $G$, it is impossible to measure either the unlensed flux $F_0$ or, more importantly, the time scale $\omega^{-1}$.

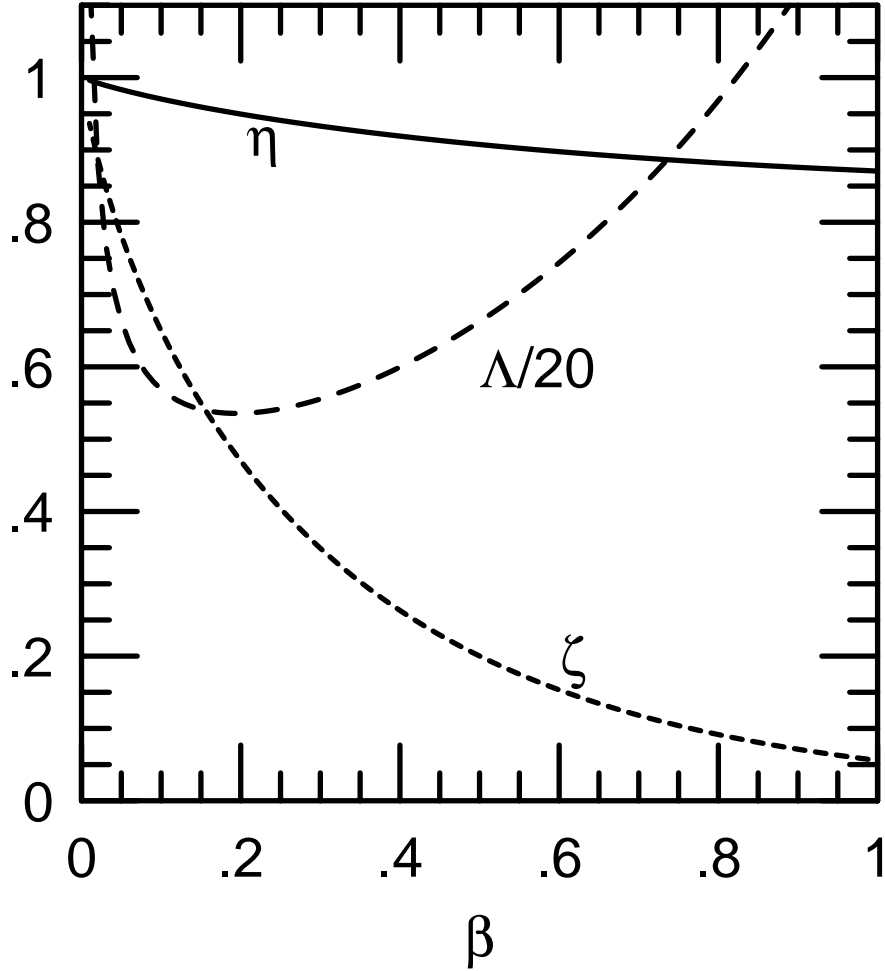

Figure 1. Three functions which help characterize the relation of the flux excess function $[A(t;t_0,\beta,\omega)-1]$ [see eq. (1.1)] to its limiting form $G(t;t_0,\omega/\beta)$ [see eq. (2.5)]. The function $\eta(\beta)$ is the correlation between $(A-1)$ and $G$. The fact that $\eta \sim 1$ shows that the $G$'s are excellent filters to search for lensing events. However, since $G$ contains no information about the lensing time scale $\omega^{-1}$, this fact also means that it is difficult to recover $\omega$ from $(A-1)$. The function $\Lambda(\beta)$ quantifies this difficulty: if the event is detected with signal to noise $Q$, then the fractional error in the time scale is $Q/\Lambda$. The function $[\zeta(\beta)]^{1/2}$ is the ratio of the true signal-to-noise for $(A-1)$ compared to what it would be if the excess flux function were actually $G$.

Thus, the first step in analyzing pixel lensing is to quantify the difference between $(A-1)$ and $G$ as a function of $\beta$. To this end I define three functions, $\eta(\beta)$, $\Lambda(\beta)$, and $\zeta(\beta)$.

Figure 1 shows $\eta(\beta)$, the correlation between $(A-1)$ and $G$,

$$\eta(\beta) = \frac{\int dt (A(t;t_0,\beta,\omega) - 1) G(t;t_0,\omega_{\text{eff}})}{\{\int dt [A(t;t_0,\beta,\omega) - 1]^2\}^{1/2} [\int dt G(t;t_0,\omega_{\text{eff}})^2]^{1/2}}. \qquad (2.6)$$

Note that $\eta(\beta) > 0.87$ over the entire range $0 \leq \beta \leq 1$. This shows that even classical lensing events are reasonably well represented by the filter function (which contains no information about the event time scale $\omega^{-1}$). It implies that even in classical lensing, the time scale is a relatively degenerate parameter. It also implies that the two-parameter functions $G$ are reasonably efficient filters to use to search for microlensing light curves, even when the events have relatively large impact parameters. See § 8.

I define $\Lambda(\beta)$ so as to quantify the precision of the time-scale measurement. Suppose a series of measurements $F(t_i)$ are made at uniform intervals which are short compared to $\omega_{\text{eff}}^{-1}$ for a duration long compared to $\omega^{-1}$. And suppose that each measurement has the same noise $\delta F$. Then the total signal to noise of the event $Q$ is given by $Q^2 = (F_0/\delta F)^2 \sum_i [A(t_i) - 1]^2$. The same measurements can also also be used to determine $\omega$, and this determination will have some error $\delta\omega$. I define $\Lambda \equiv Q\delta\omega/\omega$. That is, $\Lambda$ tells us how much more difficult it is to measure a time scale than to merely detect an event. Figure 1 displays $\Lambda/20$. The important point is that for $0.05 \lesssim \beta \lesssim 0.5$, $\Lambda$ is approximately constant at $\sim 10$. This means that an event can be detected with good confidence ($Q \sim 10$) and still yield no information about the time scale. Just to make a measurement of the time scale at $\sim 20\%$ accuracy requires $Q \gtrsim 50$.

The last quantity displayed in Figure 1,

$$\zeta(\beta) \equiv \frac{\int dt [A(t;t_0,\beta,\omega) - 1]^2}{\beta^{-2} \int dt G(t;t_0,\omega_{\text{eff}})^2}, \qquad (2.7)$$

is useful for estimating $Q$. As I will discuss in § 5, for $\beta \ll 1$, $Q^2 \propto F_0^2/\beta\omega$. The function $\zeta$ is a correction factor to this equation for finite $\beta$: $Q^2 \propto F_0^2 \zeta(\beta)/\beta\omega$. For fixed $F_0$ and $\omega$, one then finds $\delta\omega/\omega \propto \Lambda(\beta)[\beta/\zeta(\beta)]^{1/2}$. This combination is monotonic in $\beta$ which means that if the time scale can be measured to a given precision for a given $\beta$ then the measurement improves at smaller $\beta$.

## 3. Beyond the Crowding Limit

Let us suppose that a candidate lensing event is identified and that its center $\mathbf{z} = (z_1, z_2)$ on the difference image $D_{ij}(t)$ is known. Here $(i, j)$ denote the integer pixel position, while $\mathbf{z}$ is a pair of real numbers. Let $\psi_{\mathbf{z};ij}$ be the PSF centered at $\mathbf{z}$ and normalized so that $\sum_{ij} \psi_{\mathbf{z};ij} = 1$. Let the mean number of photons per pixel in the original image $I_{ij}(t)$ in the neighborhood of $\mathbf{z}$ be $n$. Then the best estimate of the number of photons in the stellar image is $\Omega_{\mathrm{psf}} \sum_{ij} \psi_{\mathbf{z};ij} D_{ij}(t)$ and the variance of the estimate is $n\Omega_{\mathrm{psf}}$, where the resolution element $\Omega_{\mathrm{psf}}$ is defined by,

$$\Omega_{\mathrm{psf}}^{-1} \equiv \sum_{ij} \psi_{\mathbf{z};ij}^2. \qquad (3.1)$$

For a Gaussian PSF, $\Omega_{\mathrm{psf}} = 4\pi\sigma^2$ where $\sigma$ is the Gaussian half-width. For Nyquist or "critically sampled" data ($\sigma = 1\,\mathrm{pixel}$), $\Omega_{\mathrm{psf}} \sim 12.6\,\mathrm{pixels}$. (I use the same symbol $\Omega_{\mathrm{psf}}$ to express the resolution element in pixels or arcsec$^2$.)

Note that $\Omega_{\mathrm{psf}}$ can vary over the image on two distinct scales. First, because of discrete pixelization, the shape of the PSF will depend on where the center is relative to the pixel structure. If the data are oversampled, then the pixelized PSF will vary smoothly and the sum in equation (3.1) will be well approximated by an integral. In this case, $\Omega_{\mathrm{psf}}$ varies very little with position. The more undersampled the data, the more $\Omega_{\mathrm{psf}}$ will vary. However, even for the significantly undersampled *Hubble Space Telescope (HST)* WFC2, for which I find a mean $\langle \Omega_{\mathrm{psf}} \rangle = 8.5\,\mathrm{pixels}$, the standard deviation around this mean is only 11%. For purposes of this paper, such small variations can be ignored. The resolution element can also vary across the field because of telescope optics. However, since all analyses presented here will be conducted locally, this variation can also be ignored.

If the density of detectable stars on an image is $\Omega_{\mathrm{psf}}^{-1}$, it is said to be "crowding limited". Below the crowding limit, an increase of the exposure time increases the number of resolvable stars as more of the luminosity function rises above the noise. Once this limit is reached, however, further increases in exposure time has only a marginal impact on the number of resolvable stars. In classical lensing studies, the maximum number of stars which can be monitored is equal to the number of resolution elements, $N_{\mathrm{res}}$,

$$N_{\mathrm{res}} = \frac{\Omega_{\mathrm{ccd}}}{\Omega_{\mathrm{psf}}}, \qquad (3.2)$$

where $\Omega_{\mathrm{ccd}}$ is the area of the CCD. Classical lensing is therefore optimized if the sampling is just critical (to maximize $N_{\mathrm{res}}$) and if the exposure times are just sufficient to reach the crowding limit.

The optimal criteria for classical lensing are not determined by the nature of the lensing events, but by the nature of the data reduction programs that are employed to detect them. These programs have their origin in the problem of doing photometry in crowded fields, such as globular clusters. The goal of such observations is to measure the fluxes from as many stars as possible. However, measuring the flux from individual stars is *completely irrelevant* in microlensing studies. *All* of the information is contained in the flux differences $F_0(A-1)$. These italicized claims seem to contradict common sense because there certainly would be additional information if one could measure the unlensed flux $F_0$. As I have emphasized, however, what is measurable in crowded fields is not $F_0$ but $B = F_0 + B'$. See equations (2.1) and (2.2).

The "crowding limit" is not a fundamental barrier to pixel lensing. Deeper exposures increase the number of stars per resolution element that are effectively monitored. In principle, this number is limited only by the base of the LF at $M_I \sim 14$. In practice, the value of obtaining deeper exposures saturates at a certain point. This point is a function of the characteristics of the galaxy (distance, surface brightness, LF) and also of the telescope. For some galaxies (like the bulge of the Milky Way) this point of diminishing returns is in fact the crowding limit. For more distant galaxies, it is well beyond the crowding limit.

## 4. The Fluctuation Magnitude Star

There are two fundamental parameters which characterize pixel lensing. The first is the resolution element $\Omega_{\rm psf}$ given by equation (3.1). The second is the "fluctuation flux" $F_*$, the ratio of the first two moments of the LF $\phi(F)$,

$$F_* \equiv \frac{\int dF\, \phi(F) F^2}{\int dF\, \phi(F) F}. \qquad (4.1)$$

The fluctuation flux (and the corresponding fluctuation magnitude) are of course also fundamental to the surface-brightness-fluctuations (SBF) method of measuring distances (Tonry 1991). For $I$ band, the absolute fluctuation magnitude is $\bar{M}_I = -4.84 + 3.0(V-I)$. The distance modulus is determined by measuring the fluctuation flux, converting to an apparent fluctuation magnitude, $\bar{m}_I$, correcting for extinction $A_I$, and then subtracting $\mu = \bar{m}_I - \bar{M}_I - A_I$. In lensing only the empirically determined $\bar{m}_I$ comes into play.

The fluctuation magnitude star enters almost every aspect of pixel lensing. As I show in the next section, the detectable event rate in the spike regime is directly proportional to the rate of photon detection from a fluctuation magnitude star.

In the following section I show that $F_*$ helps define the boundary between semi-classical and spike regimes. In § 9, I show that the relationships between photon noise and various types of systematic error can be quantified in terms of $F_*$.

## 5. Spike Pixel Lensing

Consider a galaxy with surface brightness $S$ observed when the sky brightness is $S_{\rm sky}$. Define $\kappa = (1 + S_{\rm sky}/S)^{-1}$. Let $\alpha$ be the number of photons collected by the telescope per unit time per unit flux, and let $t_{\rm ex}$ be the exposure time. The signal to noise of a measurement of excess flux due to a lensing event at time $t_i$ is then,

$$Q_i = F_0[A(t_i) - 1]\left(\frac{\alpha t_{\rm ex}\kappa_i}{S\Omega_{\rm psf}}\right)^{1/2}. \qquad (5.1)$$

(I have ignored readout noise here. However, it could easily be included in this formalism by augmenting the sky noise. For example, if the readout noise is 10 $e^-$ per pixel, the sky counts could be increased by $10^2$.)

The total signal to noise of the event is then given by $Q = (\sum_i Q_i^2)^{1/2}$. I now make three simplifying assumptions. First, I assume that the observations are carried out uniformly at intervals small compared to $\omega_{\rm eff}^{-1}$. Second, I assume that the sky noise and hence $\kappa$ is constant over time. Third, I assume that $\Omega_{\rm psf}$ is constant. The sum can then be converted into an integral which yields,

$$Q^2 = \frac{\pi F_0^2 \bar{\alpha}\kappa}{S\Omega_{\rm psf}\omega}\frac{\zeta(\beta)}{\beta}, \qquad (5.2)$$

where $\bar{\alpha}$ is the photon detection rate averaged over the duty cycle of the telescope and $\zeta$ is the suppression function given by equation (2.7) and shown in Figure 1.

The total signal to noise is a monotonic function of $\beta$, so that for fixed $F_0$ and $\omega$, and for a given threshold of detectability $Q_{\rm min}$, there is a maximum impact parameter $\beta_{\rm max}$ such that events below this value are detectable and events above it are not. Note from Figure 1 that $\zeta \to 1$ in the limit $\beta \ll 1$ and that $\zeta$ falls rapidly for $\beta \gtrsim 1/4$. For simplicity I approximate $\zeta(\beta) \sim \Theta(1/4 - \beta)$, where $\Theta$ is a step function. The maximum impact parameter is then,

$$\beta_{\rm max}(F_0) = \min\left\{\frac{\pi F_0^2 \bar{\alpha}\kappa}{Q_{\rm min}^2 \omega S\Omega_{\rm psf}}, \frac{1}{4}\right\}. \qquad (5.3)$$

The rate of detectable events from a given star is just $\beta_{\rm max}\Gamma_0$ where $\Gamma_0$ is the classical event rate given by equation (1.5). Integrating this rate over the LF and the area of the CCD, I find

$$\Gamma = \frac{2}{Q_{\min}^2} \tau N_{\text{res}} \kappa \Gamma_* \xi, \qquad \Gamma_* \equiv \bar{\alpha} F_* \qquad (5.4)$$

where $\Gamma_*$ is the photon detection rate from a fluctuation magnitude star, and $\xi$ is a correction factor arising from the the upper limit $\beta_{\max} \leq 1/4$,

$$\xi = \frac{\int_{F_{\min}}^{\infty} dF\, \phi(F)(\min\{F, F_{\max}\})^2}{\int_0^{\infty} dF\, \phi(F) F^2}, \qquad F_{\max}^2 \equiv \frac{Q_{\min}^2 \omega S \Omega_{\text{psf}}}{4\pi \bar{\alpha} \kappa}. \qquad (5.5)$$

Here $F_{\min}$ is the minimum flux limit which results from finite-source size effects. I will discuss these effects below in § 7 but for the moment ignore them.

The flux $F_{\max}$ represents the boundary between two regimes. Lensing events of stars with $F_0 > F_{\max}$ can be recognized even when they have low impact parameters ($\beta \lesssim 1/4$). The light curves associated with these events probe the entire Einstein ring and therefore tend to give good information about the timescale $\omega^{-1}$ and the stellar flux $F_0$. Stars with flux $F_0 < F_{\max}$ give rise to events that are detectable only as spikes near the center of the Einstein ring, and which therefore do not generally give information about the timescale.

If $F_{\max}$ is brighter than $F_{\lim}$, the brightest star in the LF, then $\xi \sim 1$ and pixel lensing is in the spike regime. The event rate is then related to $\tau$ through directly observable quantities, $\Gamma_*$, $N_{\text{res}}$, $\kappa$, and $Q_{\min}$. This means that the optical depth can be determined from the observations. Note that in classical lensing the optical depth determination requires measurements of $\omega_i^{-1}$ [see eq. (1.4)]. By contrast, in the spike regime of pixel lensing the optical depth is determined with essentially no information about the individual time scales of the events. Since $\beta < 1/4$, the light curve is well described by the filter functions $G(t; t_0, \omega_{\text{eff}})$ which does not depend on $\omega$. See $\eta(\beta)$ in Figure 1.

# 6. Semi-Classical Pixel Lensing

The boundary between the spike and semi-classical regimes is the point where $F_* = F_{\max}$. Explicitly,

$$\frac{\Gamma_*/\omega}{N_*} = \frac{Q_{\min}^2}{4\pi\kappa}, \qquad N_* \equiv \frac{S\Omega_{\mathrm{psf}}}{F_*}. \tag{6.1}$$

Note that the left hand side is the ratio of the number of photons collected from a fluctuation magnitude star during an Einstein crossing time to the number of fluctuation magnitude stars in a resolution element. If the exposures are much less deep than this, then $\xi \sim 1$, the case discussed in the previous section. If the exposures are much deeper than this (the semi-classical regime) then $\xi \ll 1$. Equation (5.4) remains formally valid, but is no longer a very useful way to express the event rate.

In particular, what one gains from deeper exposures is different in the semi-classical and spike regimes. For spike pixel lensing, the total number of events rises directly with $\bar{\alpha}/\Omega_{\mathrm{psf}}$. For semi-classical lensing, the event rate grows much more slowly. This is well illustrated by the simple quasi-realistic LF $\phi(F) \propto F^{-1}\Theta(F_{\lim} - F)$, for which $\xi = (F_{\max}/F_{\lim})^2[1 + 2\ln(F_{\lim}/F_{\max})]$, implying $\Gamma \propto [1 + 2\ln(F_{\lim}/F_{\max})]$. The growth of $\Gamma$ with $\bar{\alpha}/\Omega_{\mathrm{psf}}$ therefore goes from linear for $F_{\max} > F_{\lim}$ to square root at $F_{\max} \sim F_{\lim}$ to logarithmic at $F_{\max} \sim F_*$, (typically $\sim 1$ mag below $F_{\lim}$). Hence, from the standpoint of detecting as many events as possible, it would be better to use the available telescope time to look at another galaxy or perhaps another field in the same galaxy. Recall from § 2, however, that if the exposures are sufficiently deep, one can measure the time scale of individual events. The time scales can be used not only to determine $\tau$, but to measure the mass spectrum of the lenses as well (Han & Gould 1996). One can also measure the individual fluxes $F_0$ of the source stars. Microlensing is probably not the most efficient method for studying the LF of the nearest galaxies, but may be the only method for distant ones. Hence the source flux measurements could prove to be uniquely valuable.

The minimum signal to noise required to measure the flux with $\sim 20\%$ accuracy is $Q \gtrsim 50$. As I will show in § 8, the minimum threshold for mere detection is $Q_{\min} \sim 7$. Thus, in order to measure time scales for even the brightest source-star events, one must reach two mag below the tip of the LF, $F_{\max} \sim 0.15 F_{\lim}$. Typically this is $\sim 1$ mag below $F_*$. Hence, there is a roughly 2 mag interval between the purely spike and purely semi-classical regimes. Equation (6.1) defines the midpoint of this transition.

While a 2 mag interval seems modest by astrophysical standards, moving through this interval requires an improvement of a factor $\sim 50$ in telescope time, telescope size, or PSF size. To a first approximation, then, whether pixel lensing is in the semi-classical or spike regime depends primarily on the galaxy's distance and only secondarily on the observing program.

## 7. Finite Size Effects

The finite angular radius of the star $\theta_s$ fundamentally limits pixel lensing. If $\beta_{\max} < \theta_s/2\theta_e$, then the maximum magnification will be underestimated by the point source formula (1.1), and the event will not in fact be seen (Gould 1995b). This occurs when $F < F_{\min}$ where,

$$F_{\min}^2 = 2\omega t_s F_{\max}^2, \qquad (7.1)$$

and $t_s = \theta_s/\omega\theta_e$ is the crossing time of the star. Note that equation (7.1) is actually a self-consistent condition because the left hand side contains the flux of the star and the right hand side contains its radius.

One can draw a few general conclusions from equation (7.1). First, if one is attempting to detect lenses in or near the source galaxy, then $t_s \sim r_s/v$ where $r_s$ is the source radius. Near the tip of the LF, $r_s \sim 0.5$ AU. For typical stellar mass $\sim 0.3 M_\odot$ and typical galaxy or cluster internal distance $d_{\mathrm{ls}} \sim 1$–$100$ kpc, the physical Einstein radius is $d_{\mathrm{ol}}\theta_e \sim 2$–$20$ AU. That is, $F_{\max}/F_{\min} \sim 2$–$6$. Hence, for spike pixel lensing to be observable at all, $F_{\max}$ must be no more than one or two mag brighter than $F_{\lim}$. Recall from § 6, that if $F_{\lim}$ is more than one or two mag fainter than $F_{\lim}$, then the pixel lensing is semi-classical. This implies that total range of spike pixel lensing is only a few mag. An important exception to this rule occurs for Galactic lenses. In this case $\theta_e$ is typically larger by a factor $d_{\mathrm{os}}(d_{\mathrm{ls}}R_0)^{-1/2}$ where $R_0$ is the galactocentric distance, so that $F_{\min}$ is almost always negligible. This implies that it is possible to use external galaxies as sources to search for Galactic lenses, even when lenses within those galaxies are undetectable.

If the time interval between exposures is longer than $t_s$, then this interval enters equation (7.1) in place of $t_s$. In fact, for ground-based observations from a single site, the situation is considerably more complicated. Observations during a single night would be sensitive to events with effective time scales $\omega_{\mathrm{eff}}^{-1} \lesssim 4\,\mathrm{hrs}$, while observations over several days would be sensitive to those with $\omega_{\mathrm{eff}}^{-1} \gtrsim 1\,\mathrm{day}$, but the sensitivity would be poor for intervening time scales.

## 8. Searching For Events

Suppose that the images all have the same PSF and all have the same $S_{\rm sky}$. One might then imagine searching for events by forming sums of the difference images $D(t_i)$,

$$\tilde{D}(t_0, \omega_{\rm eff}) = \sum_i D(t_i) G(t_i; t_0, \omega_{\rm eff}). \tag{8.1}$$

For an event with $\beta \ll 1$, effective time scale exactly equal to $\omega_{\rm eff}$, and peak exactly at $t_0$, such an image would contain a PSF with signal to noise equal to the $Q$ of the event itself. One could search the image for point sources and all such events would appear. Even for finite $\beta$, the signal to noise would only be reduced by $[\eta(\beta)]^{1/2}$ [see Fig. 1 and eq. (2.6)], a modest correction at most. If the peak differed from $t_0$ by a time $\ll \omega_{\rm eff}^{-1}$ or the time scale differed from $\omega_{\rm eff}^{-1}$ by a small fractional amount, the signal to noise would also fall by only a small amount.

This search method is a practical one for some observing programs, especially those conducted from space. In any event, the method can be used to estimate the number of independent event searches which are being made. The overlap integral between two filters $G(t; t_0, \omega_{\rm eff})$ and $G(t; t_0', \omega_{\rm eff}')$ will be high provided $|\omega_{\rm eff}(t_0' - t_0)| \lesssim 1$ and $|\ln(\omega_{\rm eff}/\omega_{\rm eff}')| \lesssim 1$. Hence there will be $\sim N_{\rm obs} \ln N_{\rm obs}$ independent combined images $\tilde{D}(t_0, \omega_{\rm eff})$, where $N_{\rm obs}$ is the ratio of the duration of the observations to the shortest effective time scale being searched. Note that $N_{\rm obs}$ will typically be equal to the number of observations. The number of independent resolution elements on each image is $N_{\rm res}$. Hence, a total of $\sim N_{\rm res} N_{\rm obs} \ln N_{\rm obs}$ independent searches are conducted. For an ambitious program, one might have $N_{\rm res} \sim 10^6$ (for moderately oversampled data on a $2048^2$ chip) and $N_{\rm obs} \sim 1\,{\rm yr}/1\,{\rm hr} \sim 10^4$. This implies $\sim 10^{11}$ independent searches. The threshold required to avoid detection of noise-induced events is then $Q_{\rm min} \sim 7$.

A more generally useful search technique would be to form images $D'(t_i)$ by replacing each pixel with the value obtained by fitting a PSF centered at that pixel to the image. Then for each pixel $(j, k)$ and each filter function $G(t; t_0, \omega_{\rm eff})$, form the signal-to-noise ratio,

$$Q = \left[n_{jk} \sum_i \frac{G(t_i; t_0, \omega_{\rm eff})^2 \kappa_i}{\Omega_{{\rm psf},i}}\right]^{-1/2} \sum_i \frac{D'_{jk}(t_i) G(t_i; t_0, \omega_{\rm eff}) \kappa_i}{\Omega_{{\rm psf},i}}, \tag{8.2}$$

where $n_{jk}$ is the number of photons per image in pixel $(j, k)$ due to the galaxy (ignoring sky). Finally search for pixels $(j, k)$ and filters at $(t_0, \omega_{\rm eff})$ for which $Q \gtrsim Q_{\rm min}$. In practice, one would actually set the threshold slightly below $Q_{\rm min}$ and then search for the local peak in $Q$ in the neighborhood of the discrete peak at $(j, k, t_0, \omega_{\rm eff})$. It is this peak which should exceed $Q_{\rm min}$.

# 9. Noise From Systematic Effects

So far I have simply assumed that the only obstacle to detecting and measuring pixel-lensing events is photon noise, an assumption that some might politely label "optimistic". For example, I assumed that the images can be aligned perfectly with no loss of information and, perhaps more strikingly, that the PSF in the reference image $R$ is identical to the one in the current image $I(t)$. I assumed that $R$ contains no noise and that cosmic ray events (CRs) play no role. Simply to list these complicating factors is enough to convince most people that pixel lensing is hopeless. However, it is possible to evaluate each effect quantitatively and to show that they do not compete with photon noise for realistic observing programs.

## 9.1. Basic Formalism

Consider a dense star field composed of stars with angular positions $\mathbf{z}_i$ and fluxes $F_i$. The surface brightness as a function of position is then given by,

$$S_0(\mathbf{z}) = \sum_i F_i \delta(\mathbf{z} - \mathbf{z}_i). \tag{9.1}$$

As seen from Earth, the surface brightness will not appear as the discrete sources represented by the $\delta$-functions in equation (9.1), but as smeared out by the PSF $\psi$ with a full width half maximum (FWHM) $\theta_{\mathrm{see}}$ which is generally of order an arcsec. The apparent surface brightness is then given by the convolution,

$$S = \psi \circ S_0. \tag{9.2}$$

If the typical separation of the stars is much smaller than $\theta_{\mathrm{see}}$, then the stars will be unresolved. This is the regime of pixel lensing.

Let us suppose that the field is imaged twice, first at time $t_1$ when none of the stars are microlensed (the reference image $R$), and later at time $t_2$, when the $j$th star is microlensed with magnification $A$. Assume for the moment that one can align the two images photometrically and geometrically (see below). The first image is subtracted from the second, and the difference image has surface brightness $\Delta S$,

$$\Delta S(\mathbf{z}) = S(\mathbf{z}, t_2) - S(\mathbf{z}, t_1) = (\psi_2 - \psi_1) \circ S_0 + (A-1) F_j \psi_2 \circ \delta(\mathbf{z} - \mathbf{z}_j), \tag{9.3}$$

where $\psi_1$ and $\psi_2$ are the PSFs at the two epochs. If we further suppose that $\psi_1 = \psi_2$, then the first term in equation (9.3) disappears. The difference image is just an isolated point source convolved with the PSF. This is the situation I assumed in § 3 where I evaluated the photon noise $F_{\gamma\,\text{noise}}$ (assuming that $R$ is noiseless),

$$F^2_{\gamma\,\text{noise}} = \frac{S\Omega_{\text{psf}}}{\kappa\alpha t_{\text{ex}}}, \qquad \Omega^{-1}_{\text{psf}} \equiv \int d^2z[\psi(\mathbf{z})]^2. \tag{9.4}$$

Recall that $t_{\text{ex}}$ is the exposure time, $\alpha$ is the photon detection rate, and $\kappa$ is the ratio of flux generated by the galaxy itself to total flux. Note that I have redefined $\Omega_{\text{psf}}$ as a continuous integral rather than a discrete sum. I consider the effects of discrete pixelization in § 9.5 below.

9.2. NOISE FROM TIME VARIATION OF THE PSF

First, I compare the noise due to time variation of the PSF with the photon noise $F_{\gamma\,\text{noise}}$. If the two PSFs in equation (9.3) are not the same, then the first term $\Delta S^{2,1} = (\psi_2 - \psi_1) \circ S_0$ will be "bumpy".

Suppose that the difference image contains a stellar profile at a given position, $(0,0)$, and suppose that the flux in the star is measured by PSF fitting, i.e., by multiplying the PSF by the observed flux. Then the residual bumps $\Delta S^{2,1}$ will add noise to this measurement. To evaluate this noise, first consider a patch of the image at position $\mathbf{z}$ and with angular area $\Delta\Omega$ which is large enough to contain many stars, but is still small compared to the PSF. If the patch contains an excess flux $F$, then the PSF fitting procedure will mis-estimate the flux of the star at $(0,0)$, by an amount $F\langle\Delta\psi_{\mathbf{z}}|\psi\rangle / \langle\psi|\psi\rangle$, where

$$\Delta\psi_{\mathbf{z}}(\mathbf{z}') \equiv \psi_2(\mathbf{z}' - \mathbf{z}) - \psi_1(\mathbf{z}' - \mathbf{z}), \tag{9.5}$$

and where the Dirac brackets indicate an inner product,

$$\langle H|\psi\rangle \equiv \int d^2z' H(\mathbf{z}')\psi(\mathbf{z}'). \tag{9.6}$$

Note from this definition and equation (9.4) that $\Omega_{\text{psf}} = \langle\psi|\psi\rangle^{-1}$. Let us assume for the moment that the mean surface brightness $\bar{S}$ of the galaxy is generated entirely by fluctuation magnitude stars with flux $F_*$. There will be a total of $\bar{S}\Delta\Omega/F_*$ such stars in the patch, and hence the square of the error induced by the patch will be

$$\Delta(F^2_{\text{psf noise}}) = \frac{\bar{S}\Delta\Omega}{F_*}[F_*\Omega_{\text{psf}}\langle\Delta\psi_{\mathbf{z}}|\psi\rangle]^2. \tag{9.7}$$

In fact, one may show that this formula remains valid for an arbitrary luminosity function. Integrating equation (9.7) over the entire PSF and dividing by equation

(9.4) yields

$$\frac{F_{\text{psf noise}}^2}{F_{\gamma\,\text{noise}}^2} = \kappa n_* \Omega_{\text{psf}} \int d^2 z \, \langle \Delta\psi_{\mathbf{z}} | \psi \rangle^2 \,, \qquad n_* \equiv F_* \alpha t_{\text{ex}}, \tag{9.8}$$

where $n_*$ is the number of photons collected during the exposure from a fluctuation magnitude star. For two Gaussian PSFs with standard deviations $\sigma_1$ and $\sigma_2$, and $\delta\sigma = \sigma_2 - \sigma_1 \ll \sigma$, $\Omega_{\text{psf}} \int d^2 z \, \langle \Delta\psi_{\mathbf{z}} | \psi \rangle^2 \sim (\delta\sigma/\sigma)^2/4$, so that

$$\frac{F_{\text{psf noise}}^2}{F_{\gamma\,\text{noise}}^2} = \frac{\kappa n_*}{4} \left(\frac{\delta\sigma}{\sigma}\right)^2. \tag{9.9}$$

In a 1 hour $I$-band exposure of M31, a 1 m telescope would collect $n_* \sim 1500$ photons from a fluctuation magnitude star ($\bar{m}_I \sim 23.5$). Hence the PSFs must be similar to within $\sim 5\%$ if the PSF noise is not to dominate. It is possible to measure the PSF from foreground stars to much better than this accuracy. Using this PSF measurement, the reference image can be convolved to the seeing of the current image (see § 9.6).

What underlying physics permits one to eliminate noise from the PSF difference, $\Delta\psi$? The noise is generated by the SBF "bumps" in the surface brightness of the galaxy. To the extent that these bumps are convolved with different PSFs, they leave behind a noisy difference image. There are $N_*$ fluctuation magnitude stars per resolution element [cf. eq. (6.1)], hence the bumps have flux $\sim N_*^{1/2} F_*$, which falls inversely with distance. For galaxies at least as far as M31, there are always field stars that are much brighter than the bumps. The PSF can therefore always be measured more accurately than is required to remove the noise from bump residuals.

The difference image consists of an isolated point-source PSF on a flat background (much like the image of a random high-latitude field) and hence it is tempting to estimate the errors in the flux using the same procedure one would use for isolated-star photometry: first measure the variance of the background pixel counts and then multiply this by the number of pixels in $\Omega_{\text{psf}}$. The product should then be variance of the flux estimate. *However, if one applies this standard approach to the problem of PSF-generated noise, one overestimates the variance by a factor* $\sim 8$. Specifically, one finds analytically that the variance of the difference image is given by $var(S^{2,1}) = F_* \bar{S} \int d^2 z [\Delta\psi(\mathbf{z})]^2$. If this variance is then measured and naively multiplied by $\Omega_{\text{psf}}$, one obtains a result which is 8 times greater than equation (9.9). It is easy to understand the source of the overestimate. In a difference between two images with similar but not identical PSFs, stars will have a characteristic profile, e.g. "rings" or "shadowed mountains". However, because the net flux in these structures is zero, and because they are anti-correlated on scales of order the PSF, they actually induce smaller errors than would fluctuations of the same amplitude but randomly distributed. Noise from geometric alignment, photometric alignment and pixelization (see below) are also overestimated by the standard procedure.

In any event, this calculation shows that it is not straight forward to estimate the flux errors from the gross noise characteristics of the difference image. Instead, error estimates should be made from repeat photometry at short time intervals and from the scatter of periodic variable star fluxes around their mean light curves.

### 9.3. Geometric Alignment

The foregoing calculation immediately permits one to estimate the accuracy required for the geometric alignment. In this case, $\psi_2$ is the same as $\psi_1$ but offset by some amount $\delta\theta$. One then finds $\Omega_{\rm psf} \int d^2z \, \langle \Delta\psi_{\rm z}|\psi\rangle^2 \sim (\delta\theta/\sigma)^2/4$, so $(F_{\rm geom\,noise}/F_{\gamma\,\rm noise})^2 \sim \kappa n_*(\delta\theta/\sigma)^2/4$. In the above example, this implies a requirement to reduce $\delta\theta \lesssim 0.05\sigma$ or about 1/10 of a pixel for mildly oversampled data. In practice, one can do much better. The underlying physical reason is the same as above: foreground stars are much brighter than the SBF bumps, so residuals in the bumps arising from uncertainty in the image position are small.

### 9.4. Photometric Alignment

After two images $I_1$ and $I_2$ are aligned geometrically, they will still suffer some photometric misalignment due to differences in the sky level and atmospheric extinction. That is $I_2 = a + bI_1$. An error $\delta b$ in estimating the linear term in this transformation can produce noise in the difference image. (Photometry is not affected by a constant offset.) The amplitude of this noise will simply be $\delta b/b$ times the amplitude of the SBF (or "graininess") of the original image. The SBF can be evaluated by substituting $\Delta\psi \rightarrow (\delta b/b)\psi$ in the previous analysis. This then implies a ratio of alignment noise to photon noise $(F_{\rm align\,noise}/F_{\gamma\,\rm noise})^2 \sim \kappa n_*(\delta b/b)^2/2$, yielding for the above example the easily reached requirement $\delta b/b \lesssim 3\%$. For galaxies at least as far as M31, photometric alignment is most easily accomplished by rescaling the images so that their means and standard deviations are equal. The "signal" for this transformation is the flux from the entire galaxy which of course is much larger than the flux from any of the bumps that one is trying to align (Melchior 1995).

## 9.5. Pixelization Noise

Finally, there is noise induced by pixelization. If the images are aligned by shifting them using a flux-conserving algorithm (like linear interpolation) and if the excess flux $(A-1)F_j$ were measured by aperture photometry, then pixelization would not induce any noise. However, aperture photometry itself introduces an enormous amount of noise. PSF fitting minimizes photometric noise and was assumed when I analyzed the sensitivity of pixel lensing in § 5. To understand pixelization noise, suppose a star with unit flux is centered in an image at $(i,j) = (0,0)$ and suppose that another image is shifted relative to the first by $(\alpha, \beta)$ fractional pixels. The flux at a point $(x, y)$ in the neighborhood of the center of pixel $(i, j)$ can be expressed as a Taylor series $\psi(x,y) = \psi(i,j) + (x-i)\psi_x(i,j) + (y-j)\psi_y(i,j)...$ where subscripts represent derivatives. Using this expression, I find that if one uses linear interpolation on the second image to approximate the flux that would have fallen in $(i, j)$ in the first, one makes an error,

$$\Delta\psi(i,j) = -\frac{1}{2}[\alpha(1-\alpha)\psi_{xx}(i,j) + \beta(1-\beta)\psi_{yy}(i,j)], \tag{9.10}$$

Within the framework of the continuous-variable formalism already developed,

$$\Delta\psi_\mathbf{z}(\mathbf{z}') = -\frac{1}{2}[\alpha(1-\alpha)\psi_{xx}(\mathbf{z}'-\mathbf{z}) + \beta(1-\beta)\psi_{yy}(\mathbf{z}'-\mathbf{z})], \tag{9.11}$$

I will use equation (9.11) rather than equation (9.10) primarily because it is more convenient. However, I should note that equation (9.11) is actually more accurate because it satisfies exactly the flux conservation condition, $\int d^2z' \Delta\psi_\mathbf{z}(\mathbf{z}') = 0$, while equation (9.10) satsifies the analogous discrete equation only approximately. Assuming a Gaussian PSF and combining equations (9.8) and (9.11) yields,

$$\frac{F^2_{\text{pixel noise}}}{F^2_{\gamma\text{ noise}}} \sim \frac{\kappa n_*}{128\sigma^4}\{3[\alpha(1-\alpha)]^2 + 2\alpha(1-\alpha)\beta(1-\beta) + 3[\beta(1-\beta)]^2\}. \tag{9.12}$$

For an ensemble of images, $\alpha$ and $\beta$ will be uniformly distributed over the interval $(0, 1/2)$ yielding a mean square error,

$$\frac{F^2_{\text{pixel noise}}}{F^2_{\gamma\text{ noise}}} \sim \frac{\kappa n_*}{500\sigma^4}. \tag{9.13}$$

Equation (9.13) implies that there is a strong incentive for obtaining highly oversampled data. An alternative would be to reduce $n_*$ by obtaining many shorter exposures. This will be necessary in any event in many cases to avoid over-exposure.

I should note that in cases of excellent pointing (e.g. *HST* where $\alpha, \beta \lesssim 0.05$ pixels) equation (9.13) should be multiplied by $\sim 30 \langle \alpha^2 \rangle$. Equation (9.13) has important implications for follow up observations of classical lensing events which I discuss in § 10.5, below.

From the discussion at the end of § 9.2, the ratio of true noise generated by systematic effects to the naive estimate one would make based on the variance in the pixel counts is given by

$$\left(\frac{\text{true noise}}{\text{naive noise}}\right)^2 = \frac{\int d^2z \, \langle \Delta\psi_{\mathbf{z}} | \psi \rangle^2}{\int d^2z [\Delta\psi(\mathbf{z})]^2}. \tag{9.14}$$

For the PSF variation, geometric misalignment, photometric misalignment, and pixelization (and for Gaussian PSFs) this ratio has values of 1/8, 1/2, 1/2, and 1/8, respectively. Thus, all systematic effects have *less* impact on the photometry than would appear based on the measured variance in pixel counts.

## 9.6. Reference Image Construction

I have so far assumed that the noise in the reference image is negligible compared to the noise in the current image. From the standpoint of photon statistics alone, this implies that there must be many more exposures which are used to construct the reference image than are used to track the event. For events that are of order the duration of the observations this is obviously impossible. The noise level for these long events will be augmented by a factor $\sim 2$ relative to what I have assumed. However, for all but the shortest observation programs, the great majority of events will be short compared to the duration of the experiment, so that photon noise in $R$ will not generally be a limiting factor.

However, it is important to recognize that the requirements on $R$ are not trivially satisfied. For example, it would not be adequate to form a single reference image out e.g., 5 "good" images that are not affected by the event, and then subtract this from each event image. In this case, the difference images $D(t_i) = I(t_i) - R$ would *individually* be dominated by the photon noise of $I(t_i)$, but in the convolved image used to search for events $\tilde{D}(t_0, \omega_{\text{eff}}) = \sum_i G(t_i; t_0, \omega_{\text{eff}}) D(t_i)$, the noise from $R$ would still dominate. This is because the contributions from $\sum_i G(t_i) I(t_i)$ would add incoherently, while the contribution from $\sum_i G(t_i) R = R \sum_i G(t_i)$ would add coherently. Either each image $I(t_i)$ must have its *own* reference image constructed of several images not in the event and not used to reference other images, or a single reference image must be constructed from a very large number of images.

These requirements lead to two different types of solutions. The first is to group the reference images by seeing and pair each group with a single $I(t_i)$ which has somewhat worse seeing than the group. Then convolve each of the members of the group to the seeing of $I(t_i)$ and form their mean excluding $3\,\sigma$ outliers. (One could also form the median, but this has $(\pi/2)^{1/2}$ more noise without, in my view, any compensating advantages.) A second solution, applicable mainly to space-based observations where the seeing is constant, is to form a single reference image from all the images. In either case, the noise of the reference images can be reduced well below that of the $I(t_i)$.

CRs are automatically removed from $R$ by the above procedure. A more delicate problem is to remove them from the $I(t_i)$. The problem only arises when searching for events for which $\omega_{\rm eff} t_{\rm ex} \gtrsim 1$. For longer events one can remove CRs in the standard way by comparing images at neighboring times. Even for the shorter events, CRs pose no problem once a candidate event has been identified: it is easy enough to check whether a particular measurement has been corrupted by a CR. CRs are only a problem in the search for short-event candidates. By searching the $D(t_i)$ for spatially correlated $3\,\sigma$ outliers whose global structure is inconsistent with a PSF, one could probably remove essentially all CRs. However, it may be that CRs become a limiting factor for short events in some cases, particularly for long, space-based exposures.

## 10. Applications

The analysis presented here can be used to evaluate prospective pixel lensing experiments and to develop new pixel lensing ideas. Here I give a few examples.

### 10.1. Classification of Target Galaxies in $I$ Band

For simplicity I assume the fluctuation magnitude is $\bar{M}_I \sim -1.2$ in all galaxies, and that a telescope with diameter $D$ collects $10(D/1\,{\rm m})^2$ photons per second from an $I = 20$ star. Critical pixel lensing ($F_* = F_{\max}$) can be evaluated from [cf. eq. (6.1)],

$$\frac{F_*}{F_{\max}} = 2\left(\frac{d_{\rm os}}{\rm Mpc}\right)^{-2}\frac{D}{1\,{\rm m}}\left(\frac{t_{\rm obs}}{\rm hr}\right)^{1/2} S_{19}^{-1/2}\left(\frac{\theta_{\rm see}}{1''}\right)^{-1}\left(\frac{Q_{\min}}{7}\right)^{-1}(15\,{\rm day}\,\omega)^{-1/2}, \tag{10.1}$$

where $t_{\rm obs}$ is the observation time per day, and $S_{19}$ is the total surface brightness (including sky) in units of 19 mag arcsec$^{-1}$.

Thus dedicated observations with a 2 m telescope would bring the disk and all but the innermost part of the bulge of M31 well within the semi-classical regime. If a substantial fraction of the halo of M31 is composed of massive compact halo objects (MACHOs), it should be possible to measure not only their total optical depth but the individual time scales of many of the events.

Even if there are no MACHOs, there will be lensing of giants in the M31 bulge by the more numerous dwarf stars in the bulge. Because the bulge of M31 is semi-classical, it is possible to measure not just the time scale of these events, but also the unlensed flux from the giant sources. That is, M31 bulge-bulge lensing can yield information on both the mass function and the LF of the M31 bulge. Han (1996) has applied the present formalism to analyze pixel lensing of M31.

There are several other large galaxies within a few Mpc, such as M81 and Cen A. Dedicated observing programs on 4 m class telescopes with excellent seeing ($\theta_{\rm see} \sim 0.''5$) could bring these into the semi-classical regime and would allow study of their mass and luminosity functions. They are accessible in the spike regime with more modest programs which would allow measurement of their optical depths.

As I have elsewhere discussed (Gould 1995b), it would be particularly interesting to search for pixel lensing toward M87 because of the possibility of detecting intra-cluster MACHOs in the Virgo cluster. Dedicated observations with *HST* just barely bring the outer part of M87 into the semi-classical regime. Most of the galaxy would be in the spike regime and the observations would yield information about the total optical depth, but not about individual time scales nor about the LF of M87.

Coma is much richer than Virgo and would therefore be an even more interesting cluster to probe for intra-cluster MACHOs. However, the $d_{\rm os}^{-2}$ scaling of equation (10.1) together with the fact that Coma is $\sim 5$ times farther than Virgo makes such a search impossible with existing equipment.

The Milky Way bulge, and the Magellanic Clouds are all so close that even with the 2–10 minute exposures of the current searches, they are well within the semi-classical regime. This is of course to be expected since the current programs are searching for classical lensing.

10.2. SPACE-BASED VERSUS GROUND-BASED PIXEL LENSING

While it is not possible with existing equipment to survey galaxies much beyond M87 for pixel lensing, such studies may be possible in the future. For diffraction limited images obtained from space, the distance $d_{\rm os}$ at which pixel lensing can be observed scales directly with telescope diameter, $D$. To see this, note that $\Gamma_* \propto (D/d_{\rm os})^2$ and $\Omega_{\rm psf} \propto D^{-2}$. For fixed number of pixels in the detector (or fixed physical size of the galaxy in the case that the field is larger than the galaxy) $\Omega_{\rm ccd} \propto D^{-2}$. Equations (5.4), (6.1), and (7.1) are all invariant under these changes.

On the other hand, for ground-based telescopes of fixed PSF, the improvement with larger aperture is much more gradual. For galaxies contained within a single CCD frame, $N_{\rm res} \propto d_{\rm os}^{-2}$. Hence the event rate [eq. (5.4)] in the spike regime and the boundary between the two regimes [eq. (6.1)] stay constant if the aperture changes with distance as $D \propto d_{\rm os}^2$ (compared to $D \propto d_{\rm os}$ from space). While the comparison is somewhat difficult, I estimate that for pixel lensing of M87 dedicated observations on Keck would be $\sim 13\%$ as effective as dedicated observations on *HST*. Pixel lensing of distant galaxies therefore appears to require space-based observations.

Various techniques are now being developed to improve resolution from the ground including adaptive optics and optical interferometry. These techniques are probably not useful for pixel lensing which requires a wide field of view over which the PSF can be measured locally with good precision.

### 10.3. INFRARED VERSUS OPTICAL PIXEL LENSING

Based on a Baade's Window K band LF provided by G. Tiede (private communication, 1995), I estimate a fluctuation magnitude $\bar{M}_K \sim -5.7$, i.e., $\sim 4.5$ mag brighter than in $I$. I estimate $\bar{M}_H \sim -5.4$. Ignoring relative detector efficiencies, this implies that $\Gamma_*$ is $\sim 16$ times higher in $K$ than $I$, and 20 times higher in $H$. However, this substantial advantage of infrared (IR) over optical pixel lensing is undercut by a number of disadvantages. First, for ground-base observations, galaxies are well below the sky in $K$ and $H$, but are mostly above the sky in $I$. Second, in the IR it is usually necessary to chop between the object and the sky in order to properly subtract the variable sky. This may not be essential for relative photometry in $K$ band, but time-variable excitation of water lines in $H$ leads to fringes which shift their angular position with time unless sky frames taken at similar times are subtracted. While the angular scales of these fringes are many arcsec and hence much larger than the PSF, the amplitudes are of order the galaxy surface brightness. Unremoved fringes would therefore make pixel lensing impossible. Since IR arrays suffer from a number of other problems including smaller size and greater non-linear response than CCDs, ground-based IR pixel lensing does not appear promising.

However, for space-based observations, $H$ band would have substantial advantages over $I$ band assuming similar quality detectors were available. From equation (5.4), $\Gamma \propto \Gamma_* \kappa / \Omega_{\rm psf}$. For diffraction limited PSFs, $(\Gamma_*/\Omega_{\rm psf})_H / (\Gamma_*/\Omega_{\rm psf})_I \sim 5$. In the inner parts of galaxies $\kappa \simeq 1$ for both $I$ and $H$. In the outer parts, the sky in $I$ cannot be neglected, particularly for Virgo and Coma which are near the ecliptic and hence prone to scattered zodiacal light. Since this scattering is weaker in $H$, the IR also has a higher $\kappa$. IR observations from space may therefore be the best approach to pixel lensing in the future assuming that present-day problems with detectors are overcome.

## 10.4. VARIABLE STARS

Variable stars are the most important background for classical lensing searches, and this is also true of pixel lensing. For spike pixel lensing, variables pose only a minor problem because the events have flux changes corresponding to $M_I \lesssim -4.5$ with characteristic times $\omega_{\text{eff}}^{-1} \lesssim 4\,\text{days}$. Variables with these characteristics are rare. However, as pixel lensing moves into the semi-classical regime, it becomes sensitive to lower-flux events and hence to contamination by a wider class of variables. On the other hand, in this regime the light curves are more accurately determined so there is a greater basis for discriminating lensing events with their characteristic profile given by equation (2.2) from variables stars. Nevertheless, the only way to take systematic account of the variable background is to measure the frequency of various classes of variables. This can be accomplished partly by searching for variables in the galaxy, a search which grows naturally out the lensing search itself. Additional information comes from variable searches in more nearby galaxies (which can be done as part of other lensing studies). The higher signal-to-noise ratio for the same luminosity star then allows one to unambiguously identify a variable in a nearby galaxy which would be difficult or impossible to pick out in a more distant galaxy. Variable searches in the Galactic bulge thus serve as a basis for estimating the types and frequencies of variable contaminants of M31 lensing studies. M31 could play the same role for M87.

In addition, variables are of interest in their own right. All four classical lensing experiments have produced extensive variable catalogs, only a small fraction of which have been published (Udalski et al. 1994b, 1995; Alcock et al. 1995d; Grison et al. 1995; Allard 1995).

Pixel lensing studies can also be used as sensitive searches for variables. As an example consider a 3-day Cepheid with mean mag $I = 21.5$ in the inner disk of M31 at a projected position where the surface brightness of the M31 bulge is $I = 19\,\text{mag arcsec}^{-2}$. Suppose that a pixel lensing study were carried out with 1 hour observation per night on a 1 m telescope in $1''$ seeing. The mean mag of the Cepheid is $\sim 6$ times brighter than $\bar{M}_I$, but since there are $N_* \sim 200$ fluctuation magnitude stars per resolution element, the Cepheid has only $\sim 1/2$ as much flux as a typical SBF. If the Cepheid were re-observed when its flux had changed by $\sim 30\%$, the difference would amount to $\sim 15\%$ of a typical SBF, and would be extremely difficult to recognize by standard techniques. Indeed, no such short-period Cepheids have been discovered within the bulge of M31. However, on a pixel lensing difference image, this flux change would yield a $\sim 5\,\sigma$ PSF, which could be detected. More important, by searching the whole series of images with a variable-star filter, the Cepheid would show up very strongly. It would not be necessary to search with a filter exactly matched to the Cepheid light curve. Sinusoidal filters could easily pick out most periodic variables. After they were identified as variables, the forms of their light curves could be more precisely measured. The same method could be used to detect much fainter variables. In fact with a somewhat more aggressive program (4 hours per night on a 2 m telescope) one could hope to detect RR Lyraes in the bulge of M31 over the course of a season.

### 10.5. Pixel Methods Applied to Classical Lensing Follow Up

At present two groups have organized follow-up observations of events found in classical lensing searches (Pratt 1995; Sackett 1995). By obtaining more accurate photometry with shorter sampling times, they hope to detect planetary or binary companions to the principal lens (Mao & Paczyński 1991; Gould & Loeb 1992; Udalski et al. 1995b; Bennett et al. 1995) measure the proper motion of the lens (Gould 1994; Nemiroff & Wickramasinghe 1994; Witt & Mao 1994), and possibly its parallax as well (Gould 1992; Alcock et al. 1995c). These measurements often require, or at least would benefit from, photometry with $\lesssim 1\%$ accuracy which is generally considered the limit for standard techniques in crowded fields.

In contrast to the classical searches themselves, the follow-up observations do not need to cover large areas and therefore suffer no pressure toward Nyquist sampling. If pixel-lensing techniques (including linear interpolation are employed) then the limit on accuracy due to pixelization is given by equation (9.13), with $n_*$ replaced by the number of photons collected from the lensed star. Thus, these searches should be conducted with at least 5 pixels per FWHM ($\sigma \gtrsim 2\,\text{pixels}$).

**Acknowledgements**: I would like to thank K. Griest, for helpful comments and discussions. This work was supported by grant AST 94-20746 from the NSF and grant NAG5-2864 from NASA.